\RequirePackage{fix-cm}

\documentclass[smallcondensed, draft]{svjour3}
\smartqed  
\usepackage{graphicx}
\usepackage[ruled,vlined]{algorithm2e}

\newtheorem{Theorem}{\bf Theorem}[section]

\newtheorem{Lemma}{\bf Lemma}[section]

\def\done{\qed}

\newcommand{\bform}{\begin{displaymath}}
\newcommand{\eform}{\end{displaymath}}
\newcommand{\beqn}[1]{\begin{equation}\label{#1}}
\newcommand{\eeqn}{\end{equation}\smallskip}
\newcommand{\beqna}[1]{\begin{eqnarray}\label{#1}}
\newcommand{\eeqna}{\end{eqnarray}\smallskip}
\newcommand{\beqnan}{\begin{eqnarray*}}
\newcommand{\eeqnan}{\end{eqnarray*}\smallskip}

\usepackage[italian, english]{babel}

\begin{document}

\title{An ${\cal O}(m\log n)$ algorithm for the weighted stable set problem in claw-free graphs with $\alpha({G}) \le 3$}

\titlerunning{Weighted stable set problem in claw-free graphs with $\alpha({G}) \le 3$}

\author{Paolo Nobili \and Antonio Sassano}

\institute{P. Nobili \at
              Dipartimento di Ingegneria dell'Innovazione, Universit\`a del Salento, Lecce, Italy \\
           \and
           A. Sassano \at
              Dipartimento di Informatica e Sistemistica ``Antonio Ruberti'',  Universit\`a di Roma ``La Sapienza'', Roma, Italy
}

\maketitle

\begin{abstract}
In this paper we show how to solve the \emph{Maximum Weight Stable Set Problem} in a claw-free graph $G(V, E)$ with $\alpha(G) \le 3$ in time ${\cal O}(|E|\log|V|)$. More precisely, in time ${\cal O}(|E|)$ we check whether $\alpha(G) \le 3$ or produce a stable set with cardinality at least $4$; moreover, if $\alpha(G) \le 3$ we produce in time ${\cal O}(|E|\log|V|)$ a maximum stable set of $G$. This improves the bound of ${\cal O}(|E||V|)$ due to Faenza et alii (\cite{FaenzaOS14}).
\end{abstract}

\section{Introduction}

\noindent
The \emph{Maximum Weight Stable Set (MWSS) Problem} in a graph $G(V, E)$ with node-weight function $w: V \rightarrow \Re$ asks for a maximum weight subset of pairwise non-adjacent \emph{nodes}. For each graph $G(V, E)$ and subset $W \subset V$ we denote by $N(W)$ (\emph{neighborhood} of $W$) the set of nodes in $V \setminus W$ adjacent to some node in $W$. If $W = \{w\}$ we simply write $N(w)$. A \emph{clique} is a complete subgraph of $G$ induced by some set of nodes $K \subseteq V$. With a little abuse of notation we also regard the set $K$ as a clique. A \emph{claw} is a graph with four nodes $w, x, y, z$ with $w$ adjacent to $x, y, z$ and $x, y, z$ mutually non-adjacent. To highlight its structure, it is denoted as $(w: x, y, z)$. A graph $G$ with no induced claws is said to be \emph{claw-free} and has the property (\cite{Berge}) that the symmetric difference of two stable sets induces a subgraph of $G$ whose connected components are either (alternating) paths or (alternating) cycles. A subset $T \in V$ is \emph{null} (\emph{universal}) to a subset $W \subseteq V \setminus T$ if and only if $N(T) \cap W = \emptyset$ ($N(T) \cap W = W$). If $T = \{u\}$ with a little abuse of notation we say that $u$ is \emph{null} (\emph{universal}) to $W$.

\medskip\noindent
Let $G(V, E)$ be a claw-free graph. A subset $X$ of $V$ is said to be \emph{local} if there exists a node $u \in V$ such that $X \subseteq N[u]$. Observe that, by \cite{KKM00}, a local set contains ${\cal O}(\sqrt{|E|})$ nodes.

\begin{Lemma}\label{3SetsLemma}
Let $G(V, E)$ be a claw-free graph and $X, Y, Z, W \subseteq V$ four disjoint local sets (with $W$ possibly empty) such that $Z$ induces a clique in $G$ and $W$ is null to $Z$. In ${\cal O}(|E|)$ time we can either find a stable set $\{x, y, z\}$ with $x \in X$, $y \in Y$, $z \in Z$ or conclude that no such stable set exists. Moreover, if $X$ is null to $Y$ and $W$ is non-empty, in ${\cal O}(|E|)$ time we can either find a stable set $\{x, y, z, w\}$ with $x \in X$, $y \in Y$, $z \in Z$, $w \in W$ or conclude that no such stable set exists.
\end{Lemma}

\noindent
\emph{Proof}. For any node $u \in X \cup Y$ let $h(u)$ denote the cardinality of $N(u) \cap Z$. It is easy to see that we can compute $h(u)$ for all the nodes $u \in X \cup Y$ in overall time ${\cal O}(|X \cup Y||Z|) = {\cal O}(|E|)$ (recall that $X$, $Y$, and $Z$ are local sets, so their cardinality is ${\cal O}(\sqrt{|E|})$). Now let $\bar x \in X$ and $\bar y \in Y$ be any two non-adjacent nodes.

\medskip\noindent
\emph{Claim (i). There exists a node $\bar z \in Z$ such that $\{\bar x, \bar y, \bar z\}$ is a stable set if and only if $h(\bar x) + h(\bar y) < |Z|$.}

\smallskip\noindent
\emph{Proof.} In fact, if $h(\bar x) + h(\bar y) < |Z|$ then the neighborhoods of nodes $\bar x$ and $\bar y$ do not cover $Z$, so there exists some node $\bar z \in Z$ which is non-adjacent to both $\bar x$ and $\bar y$. On the other hand, assume by contradiction that $h(\bar x) + h(\bar y) \ge |Z|$ and still there exists some node $\bar z \in Z$ which is non-adjacent to both $\bar x$ and $\bar y$. Let $Z' = Z \setminus \{\bar z\}$. Since we have $|N(\bar x) \cap Z'| + |N(\bar y) \cap Z'| = h(\bar x) + h(\bar y) \ge |Z'| + 1$ there exists some node $z' \in Z'$ which is adjacent to both $\bar x$ and $\bar y$. But then $(z': \bar x, \bar y, \bar z)$ is a claw in $G$, a contradiction. The claim follows.

\noindent
\emph{End of Claim (i).}

\medskip\noindent
Now, in ${\cal O}(|E|)$ time, we can check if there exists some pair of nodes $x \in X$ and $y \in Y$ such that $x, y$ are non-adjacent and $h(x) + h(y) < |Z|$. If no such pair exists, by Claim~\emph{(i)} we can conclude that no stable set $\{x, y, z\}$ with $x \in X$, $y \in Y$, $z \in Z$ exists. If, on the other hand, there exist two non-adjacent nodes $x \in X$ and $y \in Y$ satisfying $h(x) + h(y) < |Z|$ then, in ${\cal O}(\sqrt{|E|})$ time, we can find a node $z \in Z$ which is non-adjacent to both.

\medskip\noindent
Assume now that $X$ is null to $Y$. Let $\bar w$ be any node in $W$, let $\bar X = X \setminus N(\bar w)$ and let $\bar Y = Y \setminus N(\bar w)$. Since by assumption $W$ is null to $Z$, we have that there exists a stable set $\{x, y, z, \bar w\}$ with $x \in X$, $y \in Y$, $z \in Z$, if and only if there exists a stable set $\{x, y, z\}$ with $x \in \bar X$, $y \in \bar Y$, $z \in Z$. Let $\bar x \in \bar X$ and $\bar y \in \bar Y$ be two nodes such that $h(\bar x)$ and $h(\bar y)$ are minimized. We can find such nodes in ${\cal O}(\sqrt{|E|})$ time and, by assumption, $\bar x$ and $\bar y$ are non-adjacent. By Claim~\emph{(i)} and the minimality of $h(\bar x)$ and $h(\bar y)$ there exists a stable set $\{x, y, z\}$ with $x \in \bar X$, $y \in \bar Y$, $z \in Z$ if and only if $h(\bar x) + h(\bar y) < |Z|$; moreover, if such a set exists we may assume $x \equiv \bar x$ and $y \equiv \bar y$. Hence, in ${\cal O}(\sqrt{|E|})$ time we can check whether there exists a stable set $\{x, y, z, \bar w\}$ with $x \in X$, $y \in Y$, $z \in Z$. Moreover, if the check is positive in ${\cal O}(\sqrt{|E|})$ time we can find a node $\bar z \in Z$ which is non-adjacent to $\bar x$, $\bar y$ and $\bar w$ so that $\{\bar x, \bar y, \bar z, \bar w\}$ is the sought-after stable set. It follows that in ${\cal O}(|E|)$ time we can check all the nodes in $W$ and either find a stable set $\{x, y, z, w\}$ with $x \in X$, $y \in Y$, $z \in Z$, $w \in W$ or conclude that no such stable set exists. This concludes the proof of the lemma.
\done

\begin{Theorem}\label{AlphaLessThan4}
Let $G(V, E)$ be a claw-free graph. In ${\cal O}(|E|)$ time we can construct a stable set $S$ of $G$ with $|S| = \min \{\alpha(G), 4\}$.
\end{Theorem}

\noindent
\emph{Proof}. First, observe that in ${\cal O}(|E|)$ time we can check whether $G$ is a clique (in which case any singleton $S \subseteq V$ would satisfy $|S| = \alpha(G) = 1$) or construct a stable set of cardinality $2$. In the first case we are done, so assume that $\{s, t\} \subseteq V$ is a stable set of cardinality $2$.

\smallskip\noindent
We now claim that, In ${\cal O}(|E|)$ time, we can construct a stable set of cardinality $3$ or conclude that $\alpha(G) = 2$. In fact, in ${\cal O}(|V|)$ time we can classify the nodes in $V \setminus \{s, t\}$ in four sets: \emph{(i)} the set $F(s)$ of nodes adjacent to $s$ and non-adjacent to $t$; \emph{(ii)} the set $F(t)$ of nodes adjacent to $t$ and non-adjacent to $s$; \emph{(iii)} the set $W(s, t)$ of nodes adjacent both to $s$ and to $t$; and \emph{(iv)} the set $SF$ of nodes (\emph{super-free}) non-adjacent both to $s$ and to $t$. If $SF \ne \emptyset$ then let $u$ be any node in $SF$; in this case $\{s, t, u\}$ is a stable set of cardinality $3$. Otherwise, in ${\cal O}(|E|)$ time we can check whether $F(s)$ is a clique or find a pair of non-adjacent nodes $u, v \in F(s)$. If $F(s)$ is not a clique, then $\{u, v, t\}$ is a stable set of cardinality $3$. Analogously, in ${\cal O}(|E|)$ time we can check whether $F(t)$ is a clique or find a stable set of cardinality $3$. Finally, if $SF = \emptyset$ and both $F(s)$ and $F(t)$ are cliques then, by claw-freeness, a stable set $S$ of cardinality $3$ (if any) satisfies $|S \cap F(s)| = |S \cap F(t)| = |S \cap W(s, t)| = 1$. Letting $X \equiv W(s, t)$, $Y \equiv F(s)$, $Z \equiv F(t)$ and observing that $X$, $Y$, $Z$ are local sets, by Lemma~\ref{3SetsLemma} we can, in ${\cal O}(|E|)$ time, either conclude that $\alpha(G) = 2$ or find a stable set $\{x, y, z\}$ with $x \in X$, $y \in Y$, $z \in Z$. In the first case we are done, so assume that $T = \{s, t, u\} \subseteq V$ is a stable set of cardinality $3$.

\smallskip\noindent
We now claim that, In ${\cal O}(|E|)$ time, we can construct a stable set of cardinality $4$ or conclude that $\alpha(G) = 3$. In fact, in ${\cal O}(|V|)$ time we can classify the nodes in $V \setminus T$ in seven sets: \emph{(i)} the set $F(s)$ of nodes adjacent to $s$ and non-adjacent to $t$ and to $u$; \emph{(ii)} the set $F(t)$ of nodes adjacent to $t$ and non-adjacent to $s$ and to $u$; \emph{(iii)} the set $F(u)$ of nodes adjacent to $u$ and non-adjacent to $s$ and to $t$; \emph{(iv)} the set $W(s, t)$ of nodes adjacent both to $s$ and to $t$ and non-adjacent to $u$; \emph{(v)} the set $W(s, u)$ of nodes adjacent both to $s$ and to $u$ and non-adjacent to $t$; \emph{(vi)} the set $W(t, u)$ of nodes adjacent both to $t$ and to $u$ and non-adjacent to $s$; \emph{(vii)} the set $SF$ of nodes (\emph{super-free}) non-adjacent to $s$, to $t$ and to $u$. Observe that, by claw-freeness, no node can be simultaneously adjacent to $s$, $t$ and $u$, so the above classification is complete. If $SF \ne \emptyset$ then let $w$ be any node in $SF$; in this case $S = T \cup \{w\}$ is a stable set of cardinality $4$. Otherwise, in ${\cal O}(|E|)$ time we can check whether $F(s)$ is a clique or find a pair of non-adjacent nodes $v, w \in F(s)$. If $F(s)$ is not a clique, then $\{v, w\} \cup T \setminus \{s\}$ is a stable set of cardinality $4$. Analogously, in ${\cal O}(|E|)$ time we can check whether $F(t)$ or $F(u)$ are cliques or find a stable set of cardinality $4$.

\smallskip\noindent
Finally, assume that $SF$ is empty and that $F(s)$, $F(t)$, $F(u)$ are all cliques. Observe that, by claw-freeness, the symmetric difference of $T$ and any stable set $S$ of cardinality $4$ induces a subgraph of $G$ whose connected components are either paths or cycles where the nodes in $S$ and $T$ alternates. Since $|S| > |T|$, at least one component is a path $P$ with $|P \cap S| = |P \cap T| + 1$. Since $SF = \emptyset$, the path $P$ contains at least one node of $T$. If it contains a single node of $T$, say $s$, the two nodes in $P \cap S$ belong to $F(s)$, contradicting the assumption that $F(s)$ is a clique. It follows that either \emph{(i)} $P$ contains two nodes of $T$ and $|P| = 5$ or \emph{(ii)} $T \subseteq P$ and $|P| = 7$. Hence, to check whether $G$ contains a stable set $S$ of cardinality $4$ it is sufficient to verify that there exists a path $P$ of type \emph{(i)} or \emph{(ii)}. We shall prove that such check can be done in ${\cal O}(|E|)$ time.

\medskip\noindent
\emph{Case (i).}

\smallskip\noindent
We have three different choices for the pair of nodes in $P \cap T$. Consider, without loss of generality, $P \cap T = \{s, t\}$ and let $P = (x, s, y, t, z)$. Such a path exists if and only if there exists a stable set $\{x, y, z\}$ with $x \in F(s)$, $y \in W(s, t)$, $z \in F(t)$. Let $X \equiv F(s)$, $Y \equiv W(s, t)$, $Z \equiv F(t)$. Observe that $Z$ is a clique and $X$, $Y$ are local sets, so $X, Y, Z$ satisfy the hypothesis of Lemma~\ref{3SetsLemma}. Hence we can, in ${\cal O}(|E|)$ time, either find the stable set $\{x, y, z\}$ or conclude that there exists no such stable set. In the first case, observe that $u$ is non-adjacent to $x$, $y$ and $z$, so $\{x, y, z, u\}$ is the sought-after stable set of cardinality $4$.

\medskip\noindent
\emph{Case (ii).}

\smallskip\noindent
We have three different choices for the order in which the three nodes $s, t, u$ appear in the path $P$. Consider, without loss of generality, $P = (x, s, w, t, y, u, z)$. Such a path exists if and only if there exists a stable set $\{x, y, z, w\}$ with $x \in F(s)$, $y \in W(t, u)$, $z \in F(u)$, $w \in W(s, t)$. Let $X \equiv F(s)$, $Y \equiv W(t, u)$, $Z \equiv F(u)$, $W \equiv W(s, t)$. Observe that, by claw-freeness, $X$ is null to $Y$ and $W$ is null to $Z$; moreover $Z$ is a clique and $X$, $Y$, $W$ are local sets. So $X, Y, Z, W$ satisfy the hypothesis of Lemma~\ref{3SetsLemma} and we can, in ${\cal O}(|E|)$ time, either find the stable set $\{x, y, z, w\}$ or conclude that there exists no such stable set.

\medskip\noindent
It follows that in ${\cal O}(|E|)$ time we can either construct a stable set of cardinality $4$ or conclude that $\alpha(G) = 3$. This concludes the proof of the theorem.
\done

\begin{Lemma}\label{Weighted3SetsLemma}
Let $G(V, E)$ be a claw-free graph, $w \in \Re^{V}$ a weighting of $V$ and $X, Y, Z \subseteq V$ disjoint local sets such that $Z$ induces a clique in $G$. In ${\cal O}(|E|\log|V|)$ time we can either find a maximum-weight stable set $\{x, y, z\}$ with $x \in X$, $y \in Y$, $z \in Z$ or conclude that no such stable set exists.
\end{Lemma}

\noindent
\emph{Proof}. Let $z_1, z_2, \dots, z_p$ be an ordering of the nodes in $Z$ such that $w(z_1) \ge w(z_2) \ge \dots \ge w(z_p)$. Let $Z_i$ ($i = 1, \dots, p$) denote the set $\{z_1, \dots, z_i\} \subseteq Z$. For any node $u \in X \cup Y$ and index $i \in \{1, \dots, p\}$ let $h(u, i)$ denote the cardinality of $N(u) \cap Z_i$. It is easy to see that we can compute $h(u, i)$ for all the nodes $u \in X \cup Y$ and all the indices in $\{1, \dots, p\}$ in overall time ${\cal O}(|X \cup Y||Z|) = {\cal O}(|E|)$ (recall that $X$, $Y$, and $Z$ are local sets, so their cardinality is ${\cal O}(\sqrt{|E|})$). Now let $\bar x \in X$ and $\bar y \in Y$ be any two non-adjacent nodes and let $i$ be an index in $\{1, \dots, p\}$.

\medskip\noindent
\emph{Claim (i). There exists a node $\bar z \in Z_i$ such that $\{\bar x, \bar y, \bar z\}$ is a stable set if and only if $h(\bar x, i) + h(\bar y, i) < i$.}

\smallskip\noindent
\emph{Proof.} This is a special case of Claim~\emph{(i)} in Lemma~\ref{3SetsLemma}.

\noindent
\emph{End of Claim (i).}

\medskip\noindent
Now, assume $h(\bar x, p) + h(\bar y, p) < p$ and let $k$ be the smallest index in $\{1, \dots, p\}$ such that $h(\bar x, k) + h(\bar y, k) < k$.

\medskip\noindent
\emph{Claim (ii). The set $\{\bar x, \bar y, z_k\}$ is the heaviest stable set containing $\bar x$, $\bar y$ and some node in $Z$.}

\smallskip\noindent
\emph{Proof.} Trivial consequence of Claim~\emph{(i)} and the ordering of $Z$.

\noindent
\emph{End of Claim (ii).}

\medskip\noindent
\emph{Claim (iii). If $h(\bar x, i) + h(\bar y, i) < i$ for some $i \in \{1, \dots, p\}$ then $h(\bar x, j) + h(\bar y, j) < j$ for any $j \ge i$.}

\smallskip\noindent
\emph{Proof.} If $h(\bar x, i) + h(\bar y, i) < i$, by Claim~\emph{(i)} there exists a node $\bar z \in Z_i$ which is non-adjacent to both $\bar x$ and $\bar y$. If $\bar x$ and $\bar y$ had a common neighbor $z'$ in $Z_j$ then $(z': \bar x, \bar y, \bar z)$ would be a claw in $G$, a contradiction. It follows that $h(\bar x, j) + h(\bar y, j) = |N(\bar x) \cap Z_j| + |N(\bar y) \cap Z_j| < |Z_j| = j$ and the claim follows.

\noindent
\emph{End of Claim (iii).}

\medskip\noindent
By Claim~\emph{(iii)} We can find $k$ in $\lceil \log p \rceil = {\cal O}(\log|V|)$ constant time computations, by binary search. As a consequence, by checking all the pairs of non-adjacent nodes $x \in X$ and $y \in Y$, in ${\cal O}(|E|\log|V|)$ time we can either find a maximum-weight stable set $\{x, y, z\}$ with $x \in X$, $y \in Y$, $z \in Z$ or conclude that no such stable set exists. The lemma follows.
\done

\begin{Theorem}\label{MaximumAlphaEq3}
Let $G(V, E)$ be a claw-free graph and let $w \in \Re^{V}$ be a weighting of $V$. In ${\cal O}(|E|\log|V|)$ time we can either conclude that $\alpha(G) \ge 4$ or construct a maximum-weight stable set $S$ of $G$.
\end{Theorem}

\noindent
\emph{Proof}. By Theorem~\ref{AlphaLessThan4} in ${\cal O}(|E|)$ time we can construct a stable set $S$ of $G$ with $|S| = \min(\alpha(G), 4)$. If $|S| = 4$ we are done. Otherwise, $\alpha(G) \le 3$ and, as observed in \cite{FaenzaOS14}, $|V| = {\cal O}(\sqrt{|E|})$. If $|S| = \alpha(G) \le 2$ then in ${\cal O}(|E|)$ time we can find a maximum-weight stable set. In fact, since $S$ is maximal, every node in $V$ belongs to $N[S]$, $|V| = {\cal O}(\sqrt{|E|})$ and the theorem follows. Hence, we can assume that $\alpha(G) = 3$ and that we have a stable set $T = \{s, t, u\}$. Moreover, since a maximum-weight stable set intersecting $T$ can be found in ${\cal O}(|E|)$ time, we are left with the task of finding a maximum-weight stable set in $V \setminus T$. In ${\cal O}(|V|)$ time we can classify the nodes in $V \setminus T$ in six sets: \emph{(i)} the set $F(s)$ of nodes adjacent to $s$ and non-adjacent to $t$ and to $u$; \emph{(ii)} the set $F(t)$ of nodes adjacent to $t$ and non-adjacent to $s$ and to $u$; \emph{(iii)} the set $F(u)$ of nodes adjacent to $u$ and non-adjacent to $s$ and to $t$; \emph{(iv)} the set $W(s, t)$ of nodes adjacent both to $s$ and to $t$ and non-adjacent to $u$; \emph{(v)} the set $W(s, u)$ of nodes adjacent both to $s$ and to $u$ and non-adjacent to $t$; \emph{(vi)} the set $W(t, u)$ of nodes adjacent both to $t$ and to $u$ and non-adjacent to $s$. Observe that, by claw-freeness, no node can be simultaneously adjacent to $s$, $t$ and $u$; moreover, since $\alpha(G) = 3$, no node can be simultaneously non-adjacent to $s$, $t$ and $u$, so the above classification is complete. If $F(s)$ is not a clique, let $v, w$ be two non-adjacent nodes in $F(s)$. The set $\{v, w, t, u\}$ is a stable set of cardinality $4$, contradicting the assumption that $\alpha(G) = 3$. It follows that $F(s)$ and, analogously, $F(t)$ and $F(u)$ are cliques.

\smallskip\noindent
Observe that, by claw-freeness, the symmetric difference of $T$ and any stable set $S$ of cardinality $3$ induces a subgraph $H$ of $G$ whose connected components are either paths or cycles whose nodes alternate between $S$ and $T$. It follows that we can classify the stable sets non-intersecting $T$ according to the structure of the connected components of $H$. Since $\alpha(G) = 3$, no connected component of $H$ can have an odd number of nodes. We say that $S$ is of type \emph{(i)} if $H$ is a path of length $6$; of type \emph{(ii)} if $H$ is a cycle of length $6$; of type \emph{(iii)} if $H$ contains a path of length $2$. Hence, to find a maximum-weight stable set $S$ non-intersecting $T$ it is sufficient to construct (if it exists) a maximum-weight stable set of each one of the above three types. We now prove that this construction can be done in ${\cal O}(|E|)$ time.

\medskip\noindent
\emph{Case (i).}

\smallskip\noindent
If a maximum-weight stable set $S$ of type \emph{(i)} exists, then there exists a path $P$ of length $6$ containing $S$ and $T$. We have six different choices for the order of the nodes $s, t, u$ in $P$. Consider, without loss of generality, $P = (s, x, t, y, u, z)$. The set $S = \{x, y, z\}$ with $x \in W(s, t)$, $y \in W(t, u)$, $z \in F(u)$ is the sought-after maximum-weight stable set. Let $X \equiv W(s, t)$, $Y \equiv W(t, u)$, $Z \equiv F(u)$. Observe that $Z$ is a clique and $X$, $Y$ are local sets. So $X, Y, Z$ satisfy the hypothesis of Lemma~\ref{Weighted3SetsLemma} and we can, in ${\cal O}(|E|\log|V|)$ time, either find a maximum-weight stable set $\{x, y, z\}$ with $x \in X$, $y \in Y$, $z \in Z$ or conclude that no such stable set exists.

\medskip\noindent
\emph{Case (ii).} If a maximum-weight stable set $S$ of type \emph{(ii)} exists, then there exists a cycle $C$ of length $6$ containing $S$ and $T$. Let $C = (s, a, t, b, u, c)$. The set $S = \{a, b, c\}$ with $a \in W(s, t)$, $b \in W(t, u)$, $c \in W(s, u)$ is the sought-after maximum-weight stable set.

\smallskip\noindent
Assume first that $W(t, u)$ is a clique (we can check this in ${\cal O}(|E|)$ time). Let $X \equiv W(s, t)$, $Y \equiv W(s, u)$, $Z \equiv W(t, u)$. By Lemma~\ref{Weighted3SetsLemma} we can, in ${\cal O}(|E|\log|V|)$ time, either conclude that there exists no stable set of type~\emph{(ii)} or find a maximum-weight stable set of this type.

\smallskip\noindent
Assume now that $W(t, u)$ is not a clique and let $v, v'$ be two non-adjacent nodes in $W(t, u)$. Let $Z_1 = W(s, u) \cap N(v)$ and $Z_2 = W(s, u) \cap N(v')$. Since $u$ is a common neighbor to $v$, $v'$ and any node in $W(s, u)$, by claw-freeness we have $W(s, u) \subseteq Z_1 \cup Z_2$. Moreover, since $s$ is adjacent to any node in $Z_1 \cup Z_2$ and non-adjacent to $v$ and $v'$, again by claw-freeness we have $Z_1 \cap Z_2 = \emptyset$, so $Z_1$ is null to $v'$, $Z_2$ is null to $v$ and $W(s, u)$ is the disjoint union of $Z_1$ and $Z_2$. It follows that $Z_1$ is a clique for, otherwise, $(u: p, q, v')$ would be a claw, with $p$ and $q$ any two non-adjacent nodes in $Z_1$. Analogously, also $Z_2$ is a clique. Now let $X \equiv W(s, t)$, $Y \equiv W(t, u)$ and $Z \equiv Z_1$ or $Z \equiv Z_2$. By applying Lemma~\ref{Weighted3SetsLemma} twice we can, in ${\cal O}(|E|\log|V|)$ time, either conclude that there exists no stable set of type~\emph{(ii)} or find a maximum stable set $\{a, b, c\}$ with $a \in W(s, t)$, $b \in W(t, u)$, $c \in W(s, u)$.

\medskip\noindent
\emph{Case (iii).} If a maximum-weight stable set $S$ of type \emph{(iii)} exists, then there exists a path $P$ of length $2$ containing a node in $S$ and a node in $T$. We have three different choices for the node in $P \cap T$. Consider, without loss of generality, $P = (s, z)$; let $Z = F(s)$. The connected components of the symmetric difference of $S$ and $T$ containing the nodes $t$ and $u$ are either \emph{(iii-a)} two paths $P_1$ and $P_2$ of length $2$; \emph{(iii-b)} a path $P_1$ of length $4$; or \emph{(iii-c)} a cycle $C$ of length $4$. In the first case let $P_1 = (t, x)$, $P_2 = (u, y)$ and let $X = F(t)$, $Y = F(u)$. In the second case we have two possibilities: either $t$ or $u$ is an extremum of $P_1$. Without loss of generality, assume $P_1 = (t, x, u, y)$ and let $X = W(t, u)$, $Y = F(u)$. In either case, the set $S = \{x, y, z\}$ with $x \in X$, $y \in Y$, $z \in Z$ is the sought-after maximum-weight stable set. By applying Lemma~\ref{Weighted3SetsLemma} we can, in ${\cal O}(|E|\log|V|)$ time, either conclude that there exists no stable set of types~\emph{(iii-a)} and \emph{(iii-b)} or find a maximum stable set $\{x, y, z\}$ with $x \in X$, $y \in Y$, $z \in Z$. In case~\emph{(iii-c)} let $C = (t, x, u, y)$. The nodes $x$, $y$ belong to $W(t, u)$ and the node $z$ to $F(s)$. Moreover, by claw-freeness, $F(s)$ is null to $W(t, u)$. Recall that $W(t, u)$ is a local sets, so its cardinality is ${\cal O}(\sqrt{|E|})$. It follows that the maximum-weight stable set $S = \{x, y, z\}$ can be obtained by choosing the node $z$ having maximum weight in $F(s)$ and finding in ${\cal O}(|E|)$ time the pair of non-adjacent nodes $x, y \in W(t, u)$ having maximum weight. This concludes the proof of the theorem.
\done


\bibliographystyle{spmpsci}
\bibliography{ClawFreeNS}

\end{document}